# Title: Comment on "Spatially structured photons that travel in free space slower than the speed of light"

**Authors:** Z. L. Horváth* and B. Major

**Affiliations:**

Department of Optics and Quantum Electronics, University of Szeged, Dóm tér 9, Szeged H-6720, Hungary.

*Correspondence to: Z.Horvath@physx.u-szeged.hu.

**Abstract**: D. Giovannini *et al.* (Reports, 20 February 2015, p. 857) reported that they measured spatially structured photons travelling in free space slowing down even in vacuum. Here we present a simple quantum mechanical consideration which shows that even in these cases photons travel with the speed of light (*c*), and this measurement provided experimental results on the "projection" of this velocity to the axis of symmetry/beam propagation.

**One Sentence Summary:** We show that even spatially structured photons travel at the speed of light and the measurement of D. Giovannini et al. only provided the projection of this velocity onto the axis of beam propagation.

**Main Text:**

Giovannini *et al.* (*1*) has recently reported that a single photon in Bessel or Gaussian beams is delayed compared to a photon in plane waves. The delay was interpreted as the reduction of the photon velocity even if the photon travels in vacuum. We find this conclusion and the title of the paper misleading.

Here a simple quantum mechanical picture is presented to reveal the cause of the measured delay of a single photon, which shows that the delay is not due to the decrease in the velocity of the photon. We describe the field with a scalar wave function like how it was treated in (*1*) and its supplementary material, but the consideration can be generalized to vector fields by the method described in (*2*). Using scalar approximation the state of a photon in $n^{th}$ order Bessel beam (*3*) is described by the wave function

$$\psi = \psi(r, \varphi, z) = A e^{\pm i n \varphi} e^{i k z \cos\alpha} J_n(k \sin\alpha\, r),$$

where $J_n$ is the $n^{th}$ order Bessel function of the first kind, $k$ is the wave number, $(r, \varphi, z)$ are the cylindrical coordinates and the angle α is a characteristic of the beam. It can be easily shown that the Bessel beam state is an eigenstate of the component $z$ of the momentum operator with the eigenvalue of $p_z = \hbar k \cos\alpha$ since the relation $\hat{p}_z \psi = -i\hbar \cdot \partial_z \psi = (\hbar k \cos\alpha)\psi$ holds. Furthermore, ψ obeys the Helmholtz equation, that is

$$\Delta\psi + k^2\psi = 0,$$

so one can easily conclude that ψ is also an eigenfunction of the square of the momentum operator $\hat{\mathbf{p}}^2 = -\hbar^2 \Delta$ with eigenvalue $p^2 = (\hbar k)^2$. Consequently the component $z$ and the magnitude of the momentum of the photon in a Bessel beam state are definitely known (without uncertainty) and these quantities have the values of $p_z = \hbar k \cos\alpha$ and $p = \hbar k$. This means that the direction of the momentum makes an angle α with the z axis. One can also show that the expectation value of the components x and y of the momentum is zero, that is $\langle p_x \rangle = \langle p_y \rangle = 0$. However, since $\hat{p}_\perp^2 \psi = (\hat{p}^2 - \hat{p}_z^2)\psi = (\hbar k \sin\alpha)^2 \psi$ the square of the transverse component of the momentum has a definite value of $p_\perp^2 = p_x^2 + p_y^2 = (\hbar k \sin\alpha)^2$. The results show that although we do not know definitely the direction of the photon momentum, but we do know that this direction makes an angle of α with the direction of the propagation of the beam. This means that the photon in a Bessel beam state travels a longer path compared to a photon in a plane wave propagating in the direction of axis z.

It can also be generally reasoned that the velocity of the photon is the speed of light c for the following reasons. The velocity v of a free relativistic particle is determined by the momentum p and energy E since (4)

$$v = \frac{pc^2}{E}.$$

A photon state described by a wave function obeying the Helmholtz equation has the momentum $p = \hbar k$ and the energy $E = \hbar\omega$. This relation gives for the present case

$$v = \frac{\hbar k}{\hbar\omega}c^2 = \frac{c^2}{\omega/k} = c,$$

which means that the photon travels with velocity c. So the photon travels with a velocity exactly equals to the speed of light in vacuum, but in a Bessel beam it takes a "longer path" than in a plane wave state because of the change of the direction of its momentum. So the cause of the delay measured by Giovannini *et al.* is the "increased path" of the photon and not a reduced velocity.

According to our interpretation the measurement experimentally confirms that the component z of the momentum of a single photon in a Bessel beam state possess a definite value simultaneously with the magnitude of the momentum. Generally, the magnitude of the momentum and the energy of the quantum of the field always have definite values in states obeying the Helmholtz equation. Consequently, the photon travels with a velocity that exactly equals the speed of light. For Gaussian and other types of beams the experiment above measures the expectation value of the component z of the photon momentum.

An additional note is related to the expression of the group velocity appearing in both (1) and its supplementary material. The group velocity of the Bessel beam of zeroth order is well known to be superluminal (3), and can be exactly expressed as $v_g = c/\cos\alpha$. In paraxial approximation one can arrive to the expression $v_g = c(1 + \alpha^2/2)$, differing in a sign form the formula presented in (1). The term appearing as a group velocity in (1) is again the on-axis projection of the velocity of a plane wave travelling at an angle α to axis z.

**References and Notes:**

1. D. Giovannini *et al.*, *Science* 347, 6224 pp. 857-860 (2015)

**Acknowledgments:**
We would like to thank for the valuable discussions on the topic with our colleagues J. Vinkó, A. P. Kovács, M. Benedict and A. Czirják.